\documentclass{article}

\usepackage{booktabs}
\usepackage{graphics}
\usepackage{graphicx}
\usepackage{listings}
\usepackage{nameref}
\usepackage{algorithm}
\usepackage{algpseudocode}
\algrenewcommand\algorithmicrequire{\textbf{Input:}}
\algrenewcommand\algorithmicensure{\textbf{Output:}}
\usepackage{bm}
\usepackage{amsmath}
\usepackage{amssymb}
\usepackage{caption} 
\usepackage{multirow}

\usepackage{arxiv}

\usepackage[utf8]{inputenc} 
\usepackage[T1]{fontenc}    
\usepackage{hyperref}       
\usepackage{url}            
\usepackage{booktabs}       
\usepackage{amsfonts}       
\usepackage{nicefrac}       
\usepackage{microtype}      
\usepackage{lipsum}
\usepackage{graphicx}
\graphicspath{ {./images/} }

\title{Predicting the hosts of prokaryotic viruses using GCN-based semi-supervised learning}

\author{
 Jiayu Shang \\
  Dept. of Electrical Engineering\\
  City University of Hong Kong\\
  Kowloon, Hong Kong SAR, China\\
  \texttt{jyshang2-c@my.cityu.edu.hk} \\
  \And
 Yanni Sun \\
  Dept. of Electrical Engineering\\
  City University of Hong Kong\\
  Kowloon, Hong Kong SAR, China\\
  \texttt{yannisun@cityu.edu.hk} \\
}

\begin{document}

\maketitle
\begin{abstract}
\textbf{Background:} Prokaryotic viruses, which infect bacteria and archaea, are the most abundant and diverse biological entities in the biosphere. To understand their regulatory roles in various ecosystems and to harness the potential of bacteriophages for use in therapy, more knowledge of viral-host relationships is required. High-throughput sequencing and its application to the microbiome have offered new opportunities for computational approaches for predicting which hosts particular viruses can infect. However, there are two main challenges for computational host prediction. First, the empirically known virus-host relationships are very limited. Second, although sequence similarity between viruses and their prokaryote hosts have been used as a major feature for host prediction, the alignment is either missing or ambiguous in many cases. Thus, there is still a need to improve the accuracy of host prediction.\\
\textbf{Results} In this work, we present a semi-supervised learning model, named HostG, to conduct host prediction for novel viruses. We construct a knowledge graph by utilizing both virus-virus protein similarity and virus-host DNA sequence similarity. Then graph convolutional network (GCN) is adopted to exploit viruses with or without known hosts in training to enhance the learning ability. During the GCN training, we minimize the expected calibrated error (ECE) to ensure the confidence of the predictions. We tested HostG on both simulated and real sequencing data and compared its performance with other state-of-the-art methods specifcally designed for virus host classification (VHM-net, WIsH, PHP, HoPhage, RaFAH, vHULK, and VPF-Class). \\
\textbf{Conclusion} HostG outperforms other popular methods, demonstrating the efficacy of using a GCN-based semi-supervised learning approach. A particular advantage of HostG is its ability to predict hosts from new taxa.\\
\textbf{Contact} yannisun@cityu.edu.hk\\
\end{abstract}


\newpage

\section{Background}
\label{sec:intro}
Prokaryotic viruses (shortened as viruses hereafter) play an important role in the microbial system dynamics. They regulate the ecosystem by limiting the abundance of their hosts through ongoing lytic infections. Because of the threat of antibiotic resistant pathogens, there is  resurging interest of using phages as an alternative strategy to treat bacterial infections \cite{casey2018vitro}. A fundamental step in using phages to treat bacterial infection is to identify the hosts of phages, which will provide the key knowledge of using phages as potential antibiotics \cite{torres2016evolutionary}. Besides phage therapy, identifying the hosts of the novel phages have other applications such as gene transfer search \cite{canchaya2003phage, fernandez2018phage}, disease diagnosis \cite{wang2004epitope,bazan2012phage}, and novel bacterial detection \cite{edgar2006high}. 

However, despite its importance, the identified virus-host relationship is only the tip of the iceberg. The gap between the sequenced prokaryotic viruses and the known virus-host relationship is expanding quickly. Experimental methods, such as single-cell viral tagging \cite{dvzunkova2019defining}, can determine the virus-host relationship directly from the biological experiments. However, these methods are not only expensive but also time-consuming. Even worse, few virus-host connections can be detected since less than 1\% of microbial hosts have been cultivated successfully in laboratories \cite{edwards2005viral, wawrzynczak2007global}. Thus, computational approaches for predicting the host are in great demand. 

There are three main challenges for computational prediction of the virus-host relationships. First, the known virus-host interactions are limited. One of the most widely used datasets, the VHM dataset \cite{ahlgren2017alignment} contains 1,426 viruses, which is only 37\% of the known prokaryotic viruses in RefSeq. The authors of PHP \cite{lu2021prokaryotic} added virus-host relationships till 2020 from RefSeq. Together, two datasets contain around 2,000 known virus-host relationships.  Considering that prokaryotic viruses are regarded as the most abundant biological entities, the number of known interactions is still very limited compared to the unknown.  Second, although sequence similarity between viruses and prokaryotes has been used as an important feature for host identification, not all viruses share significant sequence similarities with their host genomes. In the VHM dataset, about 54\% of viruses have no alignments with the host genomes. Therefore, sequence similarity search cannot return any prediction for these viruses. Third, the finding of broad-host-range (polyvalent) phages \cite{chibani2004phage} shows that some phages can infect many different species. This poses a potential risk for binary discriminative models \cite{wang2020network, liu2019predicting, leite2018exploration}, which are designed for predicting whether a given virus-host pair represents a true infection. In their training set, all known virus-host relationships are treated as positive samples. Then, they create all-against-all virus-host pairs and often randomly select a small subset (e.g. \textasciitilde 0.5\%) of these pairs as negative samples to create a balanced dataset.  Due to the small number of negative samples, this sub-sampling method may fail to represent the original data distribution and leads to overfitting. Also, because of the presence of polyvalent phages, some pairs in the negative set can represent true infections and thus, the learned models are not reliable.

\subsection{Related work}
\label{sec:relate}
Several attempts have been made to predict hosts for viruses based on the genomic sequences \cite{roux2021global}. They can be roughly divided into two groups: alignment-based and learning-based models. Most of the alignment-based methods utilize sequence similarity search between query contigs and reference genomes of candidate hosts (bacteria or archaea). The rationale is that some viruses will preserve the borrowed genetic fragment from hosts if this genetic element brings an evolutionary advantage \cite{edwards2016computational}.  In addition, some hosts can keep a record of phage infection in CRISPR \cite{edwards2016computational}. Specifically, spacer sequences used in CRISPR systems in the host may contain such short nucleotide sequences to prevent recurring infection \cite{achigar2017phage}. Thus, CRISPR can be used as a strong signal to identify host and BLAST \cite{johnson2008ncbi} can be employed to predict hosts for viruses according to the local similarities. However, for newly identified viruses, there are two main problems for alignment-based methods, which can lead to unreliable predictions. First, viruses can share short nucleotide sequences with hosts from different taxa. According to the VHM dataset, 45.1\% virus has multiple alignment results with prokaryote in different taxa at the order level. Alignment-based tools might assign wrong taxonomic labels to viruses due to these ambiguous alignments. Second, these alignment-based methods rely heavily on the candidate hosts reference database. If the host genome is not in the database or if some viruses do not share any regions with the database, the alignment-based approaches cannot make predictions. Another solution for host prediction is to utilize the sequence similarity between viruses. For example, VPF-Class \cite{pons2021vpf} takes advantage of Viral Protein Families (VPFs) and builds a database based on the proteins from the IMG/VR system. Then, for each input contig, VPF-Class will conduct protein family search and return a prediction based on the alignment results.

Learning-based methods are more flexible. For example, VirHostMatcher (VHM) \cite{ahlgren2017alignment} and Prokaryotic virus Host Predictor (PHP) \cite{lu2021prokaryotic} utilize \textit{k}-mer-based features for prediction. VHM employs similar oligonucleotide frequency patterns between viruses and hosts and predicts the candidate host with the smallest distance for each input virus. PHP applies a Gaussian model to learn Gaussian distributions for the known virus-host pairs and outputs the probability for each input pair. Then, it uses the pair with the highest probability to assign a label for each virus. Unlike VHM and PHP, WIsH \cite{galiez2017wish} predicts the host taxon by training a homogeneous Markov model for each potential host genome. The pre-trained Markov model calculates the likelihood of the input sequence and finally predicts the host with the highest likelihood. VHM-net \cite{wang2020network} is an improved version of the VHM algorithm. This model integrates CRISPR, score of WIsH, and BLASTN results and applies Random Markov field to generate predictions. A more recently published model, RaFAH \cite{coutinho2021rafah} uses alignment features to construct a random forest for host prediction. 

The latest machine learning methods, such as deep learning algorithms, can also be used to predict hosts for viruses. To avoid manually creating negative pairs, the host prediction task can be formulated as a multi-class classification problem, where the input set contains the virus sequences and the labels are the taxa of their hosts. For example, HoPhage \cite{tan2021hophage} and vHULK \cite{amgarten2020vhulk} use a deep learning algorithm on the alignment features. However, a common problem of these methods is that they cannot predict hosts from new taxa. For example, if the training samples only contain hosts from taxa $y_1$, $y_2$, and $y_3$, these methods cannot be easily extended to predict a new host from group $y_4$.

\subsection{Overview}
Although host prediction can be formulated as a supervised-learning problem, the massive diversity of viruses and the lack of known virus-host relationships will influence the learning ability. Thus, we propose to tackle the host prediction problem in the framework of semi-supervised learning, which can better exploit the co-related information between labeled and unlabeled samples, including the organization of proteins shared between viruses and the DNA sequence similarities between viruses and prokaryote. We compared our tools with the state-of-the-art methods specifically designed for virus host classification: VHM-net, WIsH, PHP, HoPhage, RaFAH, vHULK, and VPF-Class. We also reported the results of BLASTN to show the performance of the alignment-based model. The experimental results demonstrated that HostG outperforms other popular methods. In addition, HostG can predict hosts from new taxa.

\section{Methods}
In this work, we present a method that automatically predicts the taxonomic labels (phylum to genus) of the hosts for viral contigs. Although host taxonomy prediction can be conducted on species level or even strain level, considering both polyvalent phages and the lack of known virus-host relationships, we focus on predicting the hosts' taxonomic ranking from phylum to genus in order to deliver more reliable results.

The key component of our method is the semi-supervised learning model GCN \cite{kipf2016semi}.
GCN can flexibly model the sequence-level relationships between viruses or prokaryotes using a knowledge graph and conducts convolution using node features and the topological structure. One big difference between CNN and GCN is that each node in GCN can have a different convolutional filter/kernel depending on its connections with other nodes. The convolution is conducted on each node using its own feature and the combined feature of its  neighboring nodes. Thus, the information can be passed between the labeled samples/nodes and the unlabeled samples/nodes. In biological data analyses, there exist many topological structures such as gene-sharing network, disease-drug relationship graph, and diseases-gene relationship graph. Utilizing these relationships in GCN has led to several successful applications \cite{zitnik2018modeling, stokes2020deep, chu2021mda, shang2021bacteriophage, zhao2020deeplgp}.

In our problem of predicting hosts for viruses, we will create a knowledge graph that integrates three types of information. First, although virus receptor binding proteins play an important role in helping viruses attach to the target hosts, many other proteins are involved in the process of virus infection \cite{stone2019understanding}. Thus, viruses sharing more genes tend to infect host in the same taxonomic group. The similarity of gene sharing can be represented by edges between viruses. Second, viruses and their hosts can possess local sequence similarities, which is a feature used by many available host prediction programs. The sequence similarities can be modeled as edges between viruses and hosts in the knowledge graph. Third, the nodes in the knowledge graph can be encoded as numerical vectors using a CNN for taxonomic classification of viruses. Then, the graph convolutional layer is conducted for each node and its neighbors based on the knowledge graph. The error minimization process in training will help the model fit labeled samples and back propagate the loss to the whole graph. After training, the learned convolutional filters will then be applied to predict test samples that are connected to the knowledge graph. 

\begin{figure}[h!]
    \centering
    \includegraphics[width=0.9\linewidth]{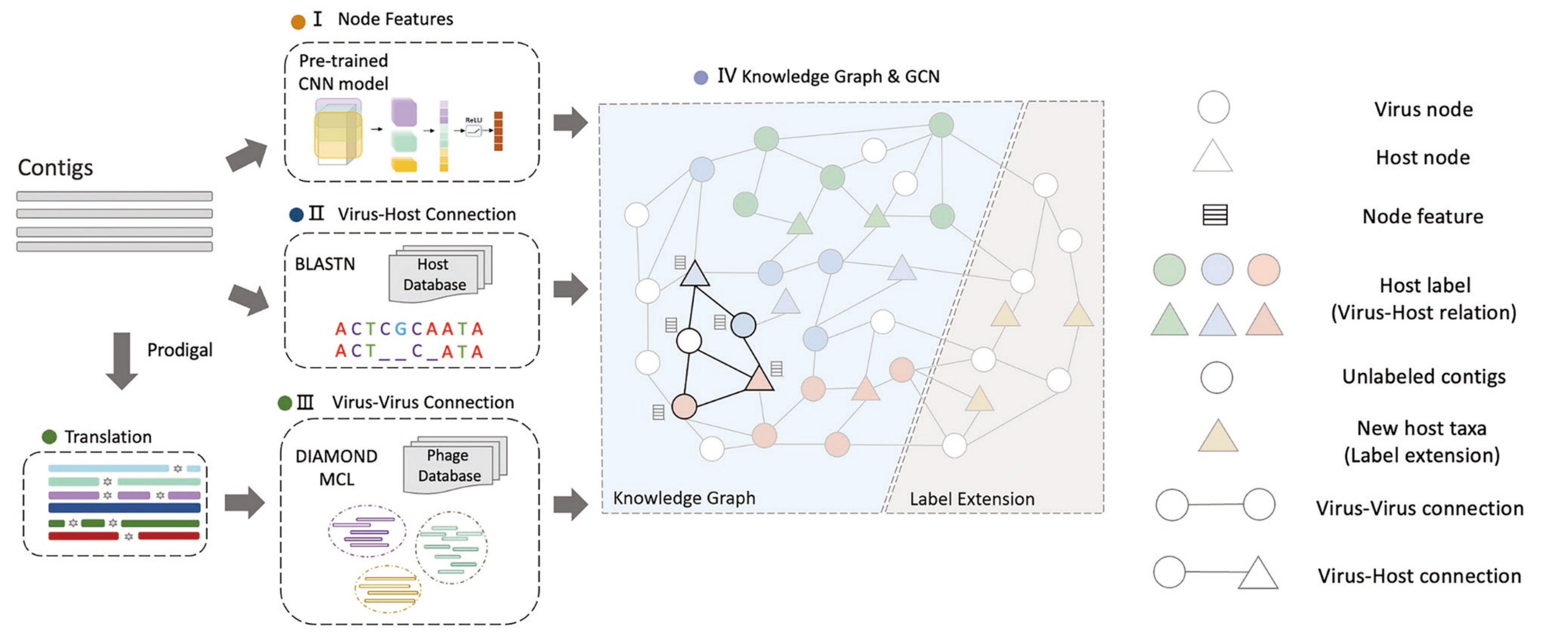}
    \caption{The pipeline of HostG. \uppercase\expandafter{\romannumeral1}: Using the pre-trained CNN model to encode contigs into node feature vectors. \uppercase\expandafter{\romannumeral2}: Utilizing BLASTN to create virus-host connections. \uppercase\expandafter{\romannumeral3}: Creating protein clusters using DIAMOND-based BLASTP and MCL. Then the protein clusters will be employed to create virus-virus connections. \uppercase\expandafter{\romannumeral4}: Creating the knowledge graph by combining the node feature and edge connections. Then GCN is employed to train and assign taxonomic labels.}
    \label{fig:figure1}
\end{figure}

\subsection{Construction of the knowledge graph}

Fig. \ref{fig:figure1} (\uppercase\expandafter{\romannumeral4}) sketches the knowledge graph. Viral and host sequences are represented by circles and triangles, respectively.  All host nodes have their taxonomic labels. For virus nodes, the colored ones are the training sequences with known hosts and thus their labels are the taxonomic labels of the hosts.  White nodes represent query virus genomes or contigs  without host information (i.e. test data). The semi-supervised learning will finally assign labels for the white nodes. 

To encode the nodes, a pre-trained CNN is applied to capture motif-related patterns from input sequence (Fig. \ref{fig:figure1}.\uppercase\expandafter{\romannumeral1}). 
There are two types of edges in the knowledge graph: virus to virus and virus to host. The edge between viruses represents sequence similarity and the similarity between shared protein families. The edge between virus and host nodes represents local similarities between the genomes.  By combining the nodes’ features and edges, we construct a knowledge graph and feed it to the GCN for training (Fig. \ref{fig:figure1}.\uppercase\expandafter{\romannumeral4}). In addition, in order to quantify the confidence of the prediction, we combine the expected calibrated error (ECE) and mean square error (L2) in the training process. After training, the knowledge graph and the learned convolution parameters are used to predict the host for new viruses. 
We will discuss the details of edge construction in Section \ref{sec:edge} and node encoding in Section \ref{sec:node}.

\subsubsection{Edge construction}
\label{sec:edge}
\paragraph{Virus-virus connection}
In this section, we first introduce the method of constructing protein clusters, which are used to establish edges between viruses (Fig. \ref{fig:figure1}.  \uppercase\expandafter{\romannumeral3}). There are three steps to construct the protein clusters. First, we extract proteins from all the virus sequences and apply DIAMOND to measure the protein similarity. For available reference genomes, the protein sequences are downloaded from NCBI RefSeq. For query/new viral contigs, we conducted gene finding and protein translation using Prodigal \cite{hyatt2010prodigal}. Then, we employ DIAMOND to conduct all-against-all pairwise alignment between contigs’ translations and reference protein sequences. DIAMOND will output the alignment of each protein pairs with E-values below a given cutoff  (the default cutoff is 1e-5). Second, based on the alignment results, we can construct a protein similarity network, where the nodes are the proteins, and the edges represent the alignments. The edge weight is the negative logarithm of the corresponding E-value. Finally, the protein clusters can be identified by the Markov clustering algorithm (MCL). 
\begin{equation}
    \label{edge1}
    P(y \ge c) =  {\textstyle \sum_{i=c}^{min(a,b)}} \frac{\binom{a}{i}\binom{n-a}{b-i}}{\binom{n}{b}}
\end{equation}
\begin{equation}
\label{edge2}
    E_{virus-virus} = \left\{\begin{matrix}
  1 ,& if \ - log(P(y \ge c) \times \binom{N}{2}) \ge \tau_1\\
   0 ,& otherwise
\end{matrix}\right.
\end{equation}

Following the idea in \cite{bolduc2017vcontact, jang2019taxonomic} , we calculate the expected number of sequences sharing at least an observed number of common proteins. By making a simplification that all protein clusters have the same probability of being chosen, we can calculate the probability of any two sequences containing $a$ and $b$ protein clusters share at least $c$ clusters by Eq. \ref{edge1}, where $y$ is the number of common protein and $n$ is the number of protein clusters. Then we calculate the expected number of sequence pairs with at least $c$ common proteins out of $\binom{N}{2}$ sequence pairs, where $N$ is the total number of sequences. As shown in Eq. \ref{edge2}, the expected value will be finally utilized to determine whether there is an edge between two sequences. The threshold $\tau_1$ is 1 by default. With the increase of $c$, $P$ become small enough to return a positive $E_{virus\raisebox{0mm}{-}virus}$. Because the size of the protein clusters varies a lot, different clusters have different probabilities of being chosen/shared. Eq. \ref{edge1} is an inaccurate but practically useful approximation in order to compute the background probability efficiently.

\paragraph{Virus-host connection}
While $E_{virus\raisebox{0mm}{-}virus}$ is used to evaluate whether two viruses share a significant number of proteins, $E_{virus\raisebox{0mm}{-}host}$ is used to measure the sequence similarity between viruses and host. We employ BLASTN to generate the sequence alignment significance between viruses and host. For $P$ virus genomes and $B$ prokaryote genomes, we will create $P\times B$ virus-host pairs. Then, as shown in Eq. \ref{edge3}, only pairs whose BLASTN E-value smaller than $\tau_2$ (default 1e-5) will form virus-host connections. Noted that if there are multiple alignment results between a virus and a potential host, we will only create one edge between them. In addition, because we have some known virus-host connections from the public dataset, we connect the viruses with their known hosts regardless of their alignment E-values.
\begin{equation}
\label{edge3}
    E_{virus\raisebox{0mm}{-}host} = \left\{\begin{matrix}
    1,& if \  virus\raisebox{0mm}{-}host\ interaction \ exists\\
  1 ,& if \ E-value <  \tau_2\\
   0 ,& otherwise
\end{matrix}\right.
\end{equation}

\subsubsection{Node construction}
\label{sec:node}
\begin{figure}[h!]
    \centering
    \includegraphics[width=0.75\linewidth]{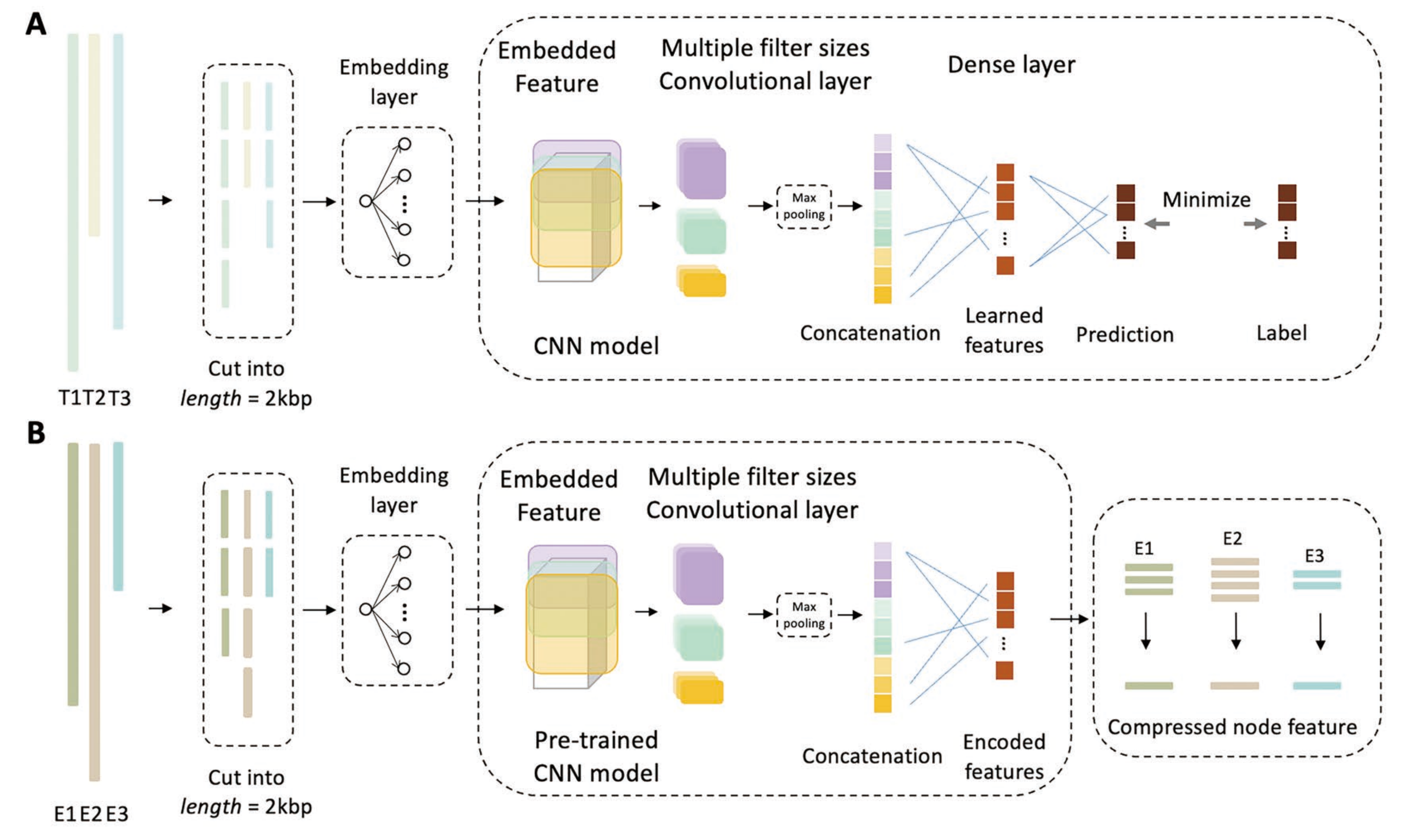}
    \caption{CNN model used in HostG. (A) Train mode. (B) Encoding mode. T1, T2, T3 represent the genomes of viruses that will be fed to CNN to update parameters during back propagation in the train mode. E1, E2, E3 represent the contigs/genomes that will be fed to pre-trained CNN for encoding the sequences into numerical vectors in the encoding mode.}
    \label{fig:figure3}
\end{figure}

Recent research shows that CNN has the ability to learn motif-related features automatically \cite{alipanahi2015predicting, du2021improving}.  Following \cite{shang2021bacteriophage}, we take advantage of CNN to learn features common to viruses of the same taxonomic group. These features are used to encode each virus node. Fig. \ref{fig:figure3} shows the training (A) and encoding (B) modes in the CNN. In the training mode, we use all the virus reference genomes with known genus labels to train the model for phage genus-level classification. In the encoding mode, the model is used to output the feature vectors of the first dense layer in the pre-trained CNN. This outputs represent encoded features of the original genomes/contigs and will be the node features in the knowledge graph.

\paragraph{CNN training}
There are four steps in the training mode. Because CNN only considers inputs with the same length, we first split the input genome into 2kbp segments. The segments have the same labels as the original genome. Second, we train a skip-gram model to convert the sequence into a numerical vector, because it can map proximate \textit{k}-mers into similar vectors in high dimensional space to inprove the learning ability \cite{mikolov2013distributed}. Thus, each 2kbp segment will be converted into a matrix $X \in \mathbb{R}^{2000-k \times d}$. $k$ is the length of \textit{k}-mers. $d$ is the number of hidden units in skip-gram model (default 100). Detailed description about the skip-gram model can be found in supplementary file.

\begin{equation}
    \label{cnn1}
    Z^{i,k}(X) =  w^k_{conv} * X[i:i+d_1-1][1:100] + b_k
\end{equation}
\begin{equation}
    \label{cnn2}
    H^{(0)} = [Maxpool(Z_0(X)),\cdots,Maxpool(Z_{N_{conv}}(X))]
\end{equation}

Third, the embedded matrix $X$ will be fed to convolutional layers. As shown in Fig. \ref{fig:figure3} (A), rather than stacking convolutional layers, we apply multiple convolutional layers with different filter sizes in parallel. With the benefit of this design, CNN can capture sequence patterns with different lengths. For each convolutional layer, the feature value at location $i$ in the $k$th feature map $Z^{i,k}(X)$ is calculated by Eq. \ref{cnn1}. $w^k_{conv}$ is the $k$th filter/kernel in the convolutional layer. $d_1$ is the filter size. $b_k$ is the bias in the $k$th feature map. Each convolutional layer will generate a high-dimensional tensor. Then, as shown in Eq. \ref{cnn2}, we apply max pooling to maintain the most important feature from these tensors and concatenate them as $H^{(0)}$. $N_{conv}$ is the total number of convolutional layers used in the structure. Detailed parameters are listed in the supplementary file.
\begin{equation}
    \label{cnn3}
    H^{(l+1)} = ReLU(H^{(l)},w^{(l)}),\ l \in \{0,1\}
\end{equation}
\begin{equation}
    \label{cnn4}
    output\ of\ train\ mode = SoftMax(H^{(2)}, w^{(2)})
\end{equation}

Finally, $H^{(0)}$ will be fed to dense layers to compress the information and make predictions. The dense layer is interpreted as Eq. \ref{cnn3}. $H^{(l)}$ is the feature map in the $l$th hidden layer. We apply the SoftMax function to generate the predictions (Eq. \ref{cnn4}) and minimize the error between labels and predictions accordingly.

\paragraph{Virus nodes}
In encoding mode, we use the pre-trained CNN to encode viruses. We use genomes/contigs as input to feed the pre-trained CNN. If the input genome is longer than 2kbp, we follow the first step in the training mode and cut it into several segments of 2kbp. As shown in Eq. \ref{cnn5}, we use the output vector of the first dense layer as the learned encoded features. Thus, the pre-trained CNN will output an encoded vector for each segment. We will add the vectors of all segments and divide the summed vector by the number of segments. 
\begin{equation}
    \label{cnn5}
    output\ of\ encoding\ mode = ReLU(H^{(0)}, w^{(0)})
\end{equation}
\paragraph{Host nodes} In order to ensure consistency in node encoding, we use weighted averaged feature vectors of viruses to encode host nodes. A virus-host connection with smaller a E-value indicates higher similarity, and thus being assigned with a bigger weight. 
\begin{equation}
    \label{cnn6}
    Node_{host} = \frac{ {\textstyle \sum_{i}^{N_{virus}}}(i) * (-log(E_{value}(i)))}{\textstyle \sum_{i}^{N_{virus}}(-log{E_{value}(i)))}} 
\end{equation}
As shown in Eq. \ref{cnn6}, the feature vector of the host node is calculated by its' neighboring virus nodes in the knowledge graph. Thus, host genomes from new taxa can still be encoded as feature vectors. This encoding method allows convenient extension of the knowledge graph to include new labels as discussed in Section \ref{sec:extend}.

\subsection{The GCN model}

\begin{figure}[h!]
    \centering
    \includegraphics[width=0.75\linewidth]{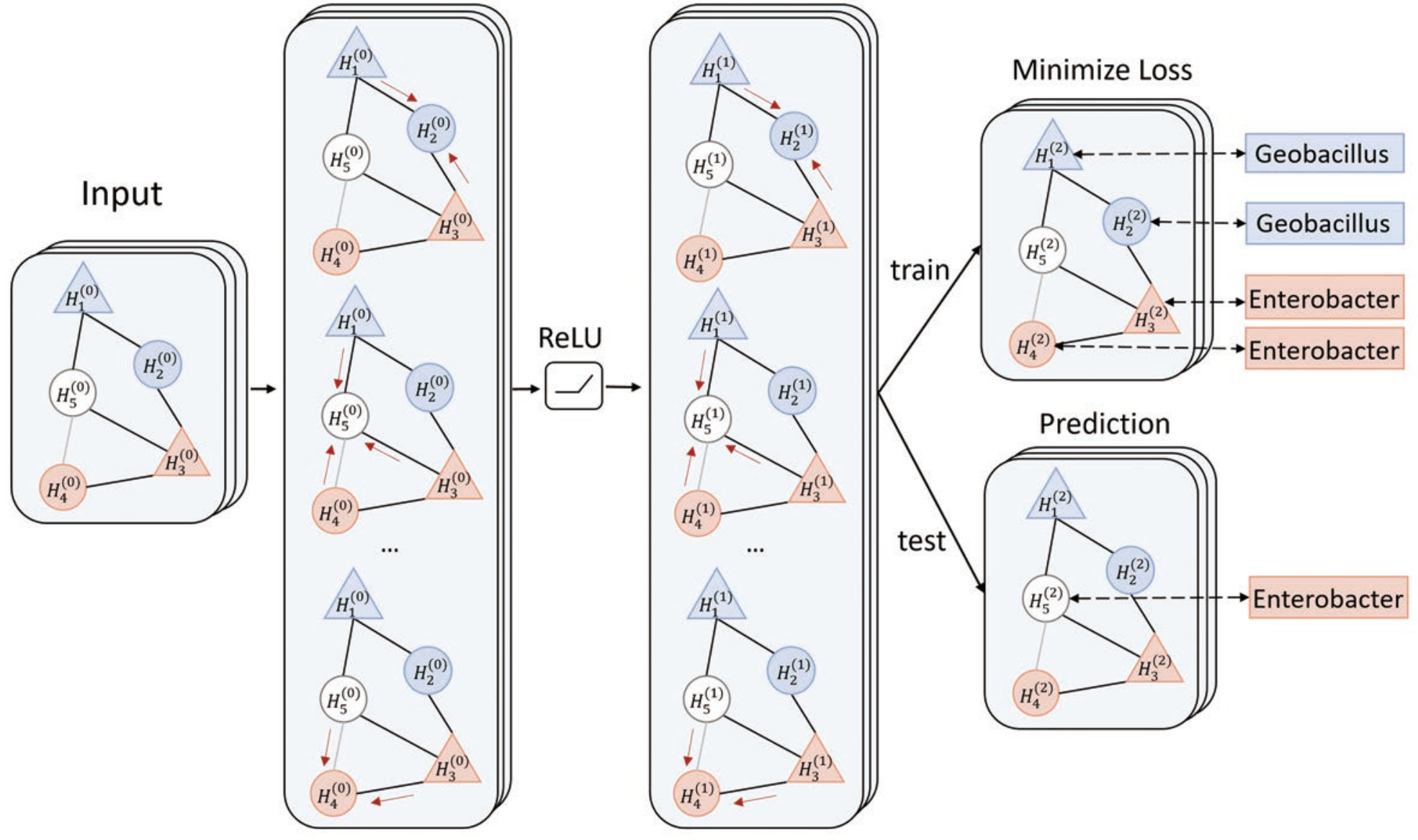}
    \caption{An example of the GCN structure in HostG. Circles represent virus nodes and triangles represent host nodes. Red color represent genus \textit{Enterobacter} and blue color represent genus \textit{Geobacillus}. White color represent query viruses. Arrows in each layer represent the graph convolution process in each layer. In the train mode, only labeled nodes will be used to minimize the loss. In the test mode, GCN will predict labels for the query node.}
    \label{fig:figure4}
\end{figure}

After constructing the knowledge graph, we train a GCN to decide the taxonomic group of the viruses’ hosts. As shown in Fig. \ref{fig:figure4}, in the graph convolutional layer, each node will use the information traversed from its’ neighbor. Eq. \ref{gcn1} shows the basic concept of the graph convolutional layer.

\begin{equation}
    \label{gcn1}
    H^{(l+1)} = ReLU (\tilde{D}^{-\frac{1}{2}} \tilde{A} \tilde{D}^{\frac{1}{2}}H^{(l)}\theta^{(l)}),\ l \in \{0,1\}
\end{equation}
\begin{equation}
    \label{gcn2}
    Out = SoftMax(H^{(2)} \theta_{dense})
\end{equation}

$A \in \mathbb{R}^{K \times K}$ is the adjacency matrix, where $K$ is the number of nodes in the knowledge graph. $\tilde{A} = A + I$, where $I \in \mathbb{R}^{K \times K}$ is the identity matrix. $\tilde{D}$ is the diagonal matrix calculated by $D_{ij} = {\textstyle \sum_{j}} \tilde{A}_{ij}$. $H^{(l)}$ is the hidden feature in the $l$th layer and $H^{(0)} \in \mathbb{R}^{K \times 512}$ is the node feature vector.  $\theta^{(l)}$ is a matrix of the trainable filter parameters in the layer. Then we feed the output of the graph convolutional layer to a dense layer and utilize the SoftMax function to give the final output (Eq. \ref{gcn2}). $\theta_{dense}$ is the weight parameters in the dense layer. During training, we calculate the SoftMax value for all nodes in the knowledge graph. Only the SoftMax value of labeled nodes will be utilized to calculate the loss and update parameters. Because all host nodes are labeled, and thus, they are also used to minimize the loss. After training, the SoftMax value of each unlabeled node will be used to assign taxonomic label accordingly. 

Since we have host prediction at multiple taxonomic rankings,  we will train one model for each taxonomic level (from genus to phylum) separately. Specifically, we re-use the same knowledge graph for training, but the label of the nodes are different according to the taxonomic level. Since there might exist inconsistencies in predicted host taxonomy, HostG will only output the higher taxonomic label when a conflict occurs.

\subsubsection{Expected calibrated error (ECE)}
\label{sec:ece}

\begin{figure}[h!]
    \centering
    \includegraphics[width=0.8\linewidth]{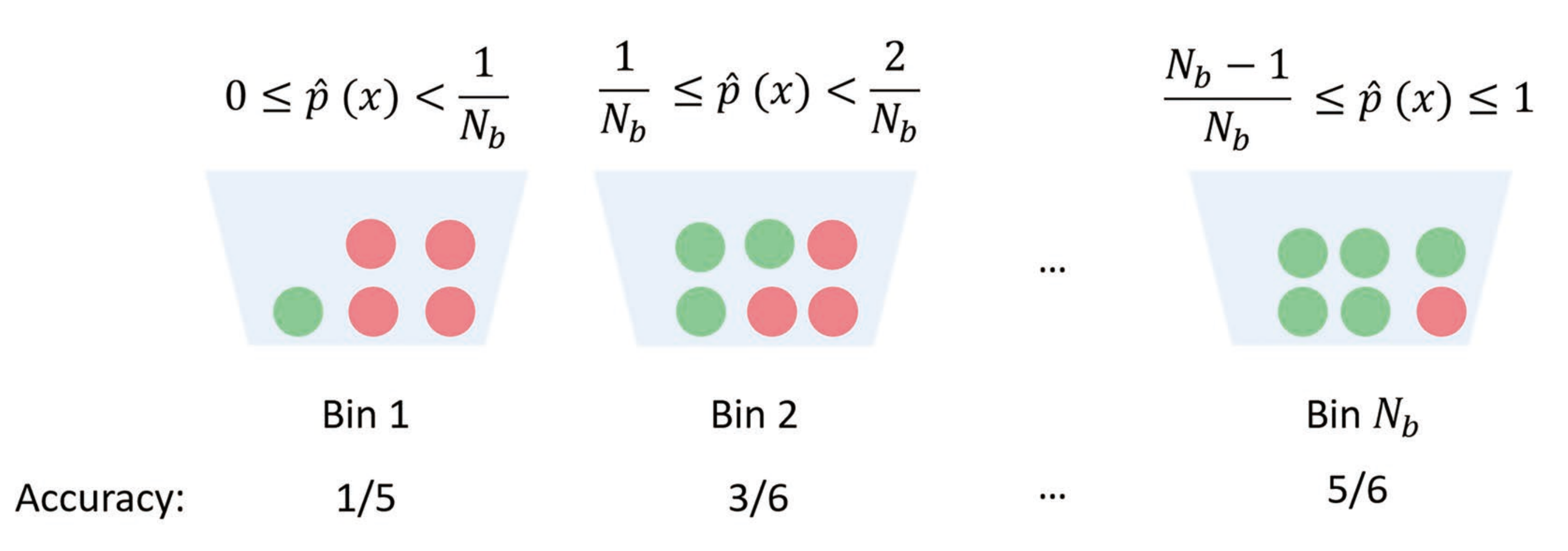}
    \caption{Example of ECE. Red circles are wrong predictions and green circles are correct predictions. $\hat{p}(x)$ is the SoftMax value of sample $x$. $N_b$ is the total number of bins.}
    \label{fig:figure5}
\end{figure}

Recent research shows that the SoftMax value cannot represent the real confidence of the prediction \cite{szegedy2013intriguing}. To improve the prediction reliability of our method, we add ECE \cite{guo2017calibration} to the objective function and update parameters with the L2 as shown in Eq. \ref{ece1}.
\begin{equation}
    \label{ece1}
    \mathcal{L} = ECE + L2
\end{equation}
ECE aims to minimize the differences between the SoftMax value and the accuracy. By updating parameters with ECE, the prediction with a higher SoftMax value will have a higher probability to be correct so that we can use the SoftMax value to represent the confidence of the prediction. We first define the ECE function. Suppose we split the SoftMax value (ranging from 0 to 1) into $N_b$ bins with each bin covering a region of size $\frac{1}{N_b}$, such as [0, $\frac{1}{N_b}$), [$\frac{1}{N_b}$, $\frac{2}{N_b}$), etc. As shown in Fig. \ref{fig:figure5}. In each training epoch, the model will output a prediction for each sample with the corresponding SoftMax value. Then we can calculate the accuracy and the average SoftMax value for each bin. Finally, as shown in Eq. \ref{ece2}, ECE is computed by the weighted sum of the difference between accuracy and average confidence (SoftMax value) in each bin. $T$ is the number of total samples and $T_i$ is the number of samples in the $i$th bin. $Acc_i$ is the accuracy of  the $i$th bin. The average confidence in each bin can be computed by Eq. \ref{ece3}. $\hat{p}(x_{ij})$ is the SoftMax value of the $j$th sample in the $i$th bin.
\begin{equation}
    \label{ece2}
    ECE =  {\textstyle \sum_{i}^{N_b}\frac{T_i}{T} \left | Acc_i - conf_i \right | } 
\end{equation}
\begin{equation}
    \label{ece3}
    conf_i = \frac{ {\textstyle \sum_{j}^{T_i}} \hat{p}(x_{ij}) } {S_i}
\end{equation}

Then we calculate the total loss of the current training epoch according to Eq. \ref{ece1} and update trainable parameters in GCN. After training with ECE loss, the difference between accuracy and average SoftMax value in each bin will become smaller. The bin with a higher SoftMax value achieves higher accuracy and thus, the SoftMax values can be used to represent the confidence of the predictions.

\subsection{Extension to new labels}
\label{sec:extend}
As the number of sequenced viruses or prokaryotes is still increasing rapidly every year, there might exist viruses infecting prokaryotes whose taxa are not included in current virus-host databases. Many existing tools can only predict the host whose taxa are in the training data. For example, if the training sequences only contain hosts from taxa $y_1$, $y_2$, and $y_3$, These tools can only learn to predict hosts with the three taxa. When the input viruses infect hosts from $y_4$, these tools are unable to give a correct prediction.

Our semi-supervised learning model allows HostG to extend to new host taxa by adding new nodes to the knowledge graph. The main idea of graph extension is to integrate new taxonomic labels by adding more host nodes into the knowledge graph. Because all the labeled nodes (including some virus nodes and all host nodes) will be used to calculate the loss and update parameters when training, these new labels will be propagated through the topological structure. Therefore, HostG not only learns from the existing virus-host interactions, but also learns the similarity between viruses and new prokaryotes to predict a new host taxa. An example is sketched in Fig. \ref{fig:figure1} (\uppercase\expandafter{\romannumeral4}). When we need to extend the GCN to include new hosts taxa (orange nodes) that do not exist in the given dataset, we will create nodes to represent these hosts. Edges connecting to the new nodes can be constructed conveniently according to $E_{virus\raisebox{0mm}{-}virus}$ and $E_{virus\raisebox{0mm}{-}host}$ described in Section \ref{sec:edge}. Then GCN will learn and traverse the information within the knowledge graph and the new taxonomic label can also be propagated by the edges. In this case, HostG can predict the new taxa for query viruses. We will demonstrate that after adding the new taxa, the model can still achieve reliable performance in the experiments. Thus, users can conveniently extend HostG to any taxa according to their needs.

\section{Result}
\label{sec:exp}
\subsection{Data and performance metrics}

\begin{table}[h!]
\centering
\begin{tabular}{cccccc}
\hline
\multicolumn{2}{c|}{The VHM dataset}              & \multicolumn{2}{c|}{The TEST dataset}           & \multicolumn{2}{c}{Single cell tagging dataset} \\ \hline
\multicolumn{2}{c|}{1,426 virus-host interaction} & \multicolumn{2}{c|}{671 virus-host interaction} & \multicolumn{2}{c}{139 virus-host interaction}  \\ \hline
\multicolumn{6}{c}{Labels of the hosts (taxonomic rank)}                                                                                              \\ \hline
Phylum          & \multicolumn{1}{c|}{7}          & Phylum         & \multicolumn{1}{c|}{7}         & Phylum                   & 2                    \\
Class           & \multicolumn{1}{c|}{13}         & Class          & \multicolumn{1}{c|}{13}        & Class                    & 4                    \\
Order           & \multicolumn{1}{c|}{36}         & Order          & \multicolumn{1}{c|}{29}        & Order                    & 4                    \\
Family          & \multicolumn{1}{c|}{67}         & Family         & \multicolumn{1}{c|}{48}        & Family                   & 5                    \\
Genus           & \multicolumn{1}{c|}{113}        & Genus          & \multicolumn{1}{c|}{64}        & Genus                    & 12                   \\ \hline
\end{tabular}
\caption{Virus-host interactions in three datasets.}
\label{tab:dataset}
\end{table}

We benchmarked our tool against other recently published host prediction tools on three datasets. The first one is the VHM benchmark dataset \cite{ahlgren2017alignment}. The taxa of both viruses and host in the dataset come from the International Committee on Taxonomy of Viruses (ICTV)  and NCBI Taxonomy database. There are 1,426 virus-host relationships in the dataset, compiled from the NCBI RefSeq before 2015. Within the 1,426 relationships, 48 viruses infect archaea and 1,378 viruses infect bacteria. The second dataset is a benchmark dataset from PHP \cite{lu2021prokaryotic}, which contains 671 virus-host interactions submitted between 2015 and 2020 (referred to as the TEST dataset hereafter). Within the 671 interactions, 21 viruses infect archaea and 650 viruses infect bacteria. The third dataset was recently constructed using single-cell viral tagging \cite{dvzunkova2019defining}. The authors identified 139 pairs of virus-host interactions.  The hosts of the three datasets come from many different taxonomic groups as shown in Table \ref{tab:dataset}. We will show the prediction performance at each taxonomic level accordingly.

\subsubsection{Experiment design}
We compared our tools with several state-of-the-art tools: WIsH \cite{galiez2017wish}, PHP \cite{lu2021prokaryotic}, HoPhage \cite{tan2021hophage}, VPF-Class \cite{pons2021vpf}, VHM-net \cite{wang2020network}, vHULK \cite{amgarten2020vhulk}, and RaFAH \cite{coutinho2021rafah}. We also recorded the output of BLASTN to show the performance of the alignment-based tool. To compare HostG with other tools fairly, we followed their experiment design and also used the same metrics: prediction rate and accuracy. These tools may return a null prediction for some samples. For example, BLASTN cannot predict the host for a virus if the viral sequence cannot be aligned with any prokaryotic genomes in the database. Thus, prediction rate is used to quantify the percentage of predicted samples as shown in Eq. \ref{m1}. It is worth noting that the prediction rate is used as recall in the benchmarked tools even though some of the predictions are not correct. Eq. \ref{m2} shows the formula to calculate the accuracy, which is computed only for samples with predicted labels. 

\begin{equation}
    \label{m1}
    prediction\ rate = \frac{number\ of\ predicted\ samples}{number\ of\ total\ samples}  
\end{equation}
\begin{equation}
    \label{m2}
    Accuracy = \frac{number\ of\ correct\ predictions}{number\ of\ predicted\ samples}
\end{equation}

The experimental results are organized as follows. First, we show the comparison between HostG and other tools at each taxonomic level. We also show that virus-virus connections can help host prediction when there is no significant alignment between viruses and host genomes. Second, we show that after combining the ECE with the L2 in GCN, the SoftMax value can represent the confidence of the prediction. Users can achieve higher accuracy with a little sacrifice of the prediction rate by specifying a confidence threshold. Third, we test how contigs of different lengths influence the performance of host prediction. Finally, we evaluate the extension ability of HostG on detecting hosts of new taxa on the second real sequencing dataset. We also compared the running time of different tools


\subsection{The performance of host taxonomy prediction}
\label{sec:exp1}
\begin{figure}[h!]
    \centering
    \includegraphics[width=0.65\linewidth]{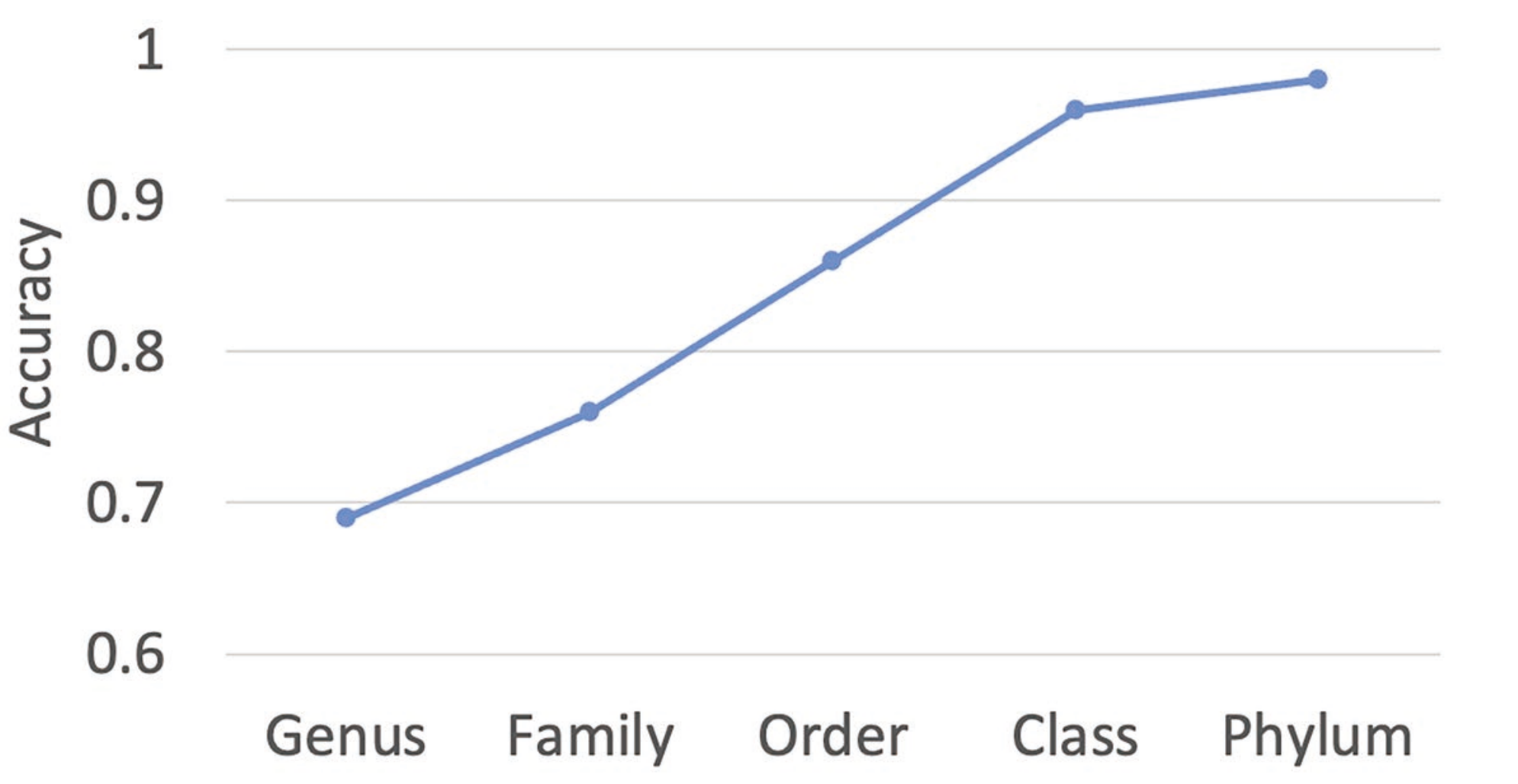}
    \caption{Average accuracy of 10-fold cross validation on VHM dataset. X-axis: taxonomic rankings. Y-axis: accuracy.}
    \label{fig:cross_valid}
\end{figure}

\begin{figure}[h!]
    \centering
    \includegraphics[width=0.8\linewidth]{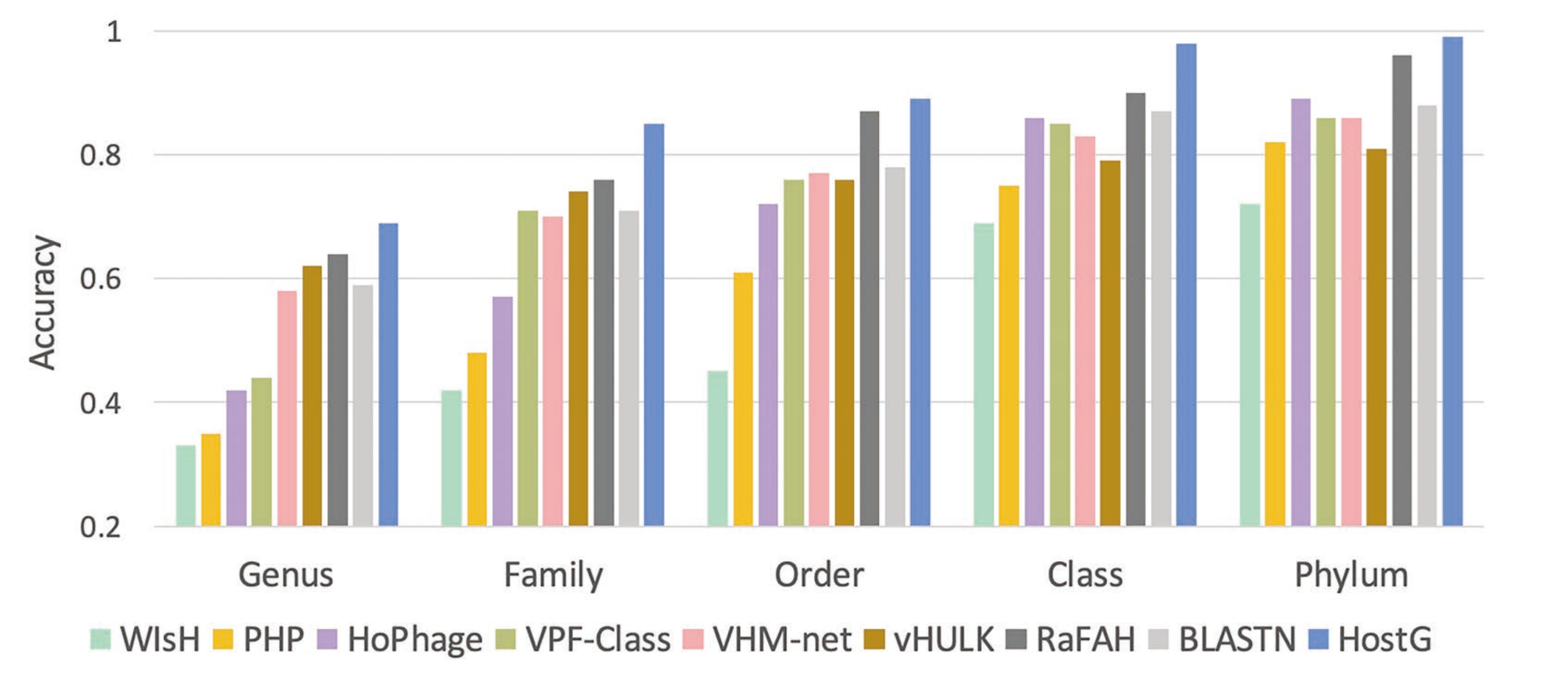}
    \caption{Host prediction accuracy from genus to phylum on the TEST dataset. X-axis: taxonomic rankings. Y-axis: accuracy.}
    \label{fig:figure6}
\end{figure}

We trained our model on the VHM dataset (interactions before 2015) and tested it on the TEST dataset (interactions between 2015 and 2020). The model parameters are learned using 10-fold cross validation on the VHM dataset. Fig. \ref{fig:cross_valid} shows the average accuracy of the 10-fold cross validation. The detailed training method of the 10-fold cross validation on the VHM dataset can be found in the supplementary file.  We also recorded the BLASTN results with the E-value cutoff 1e-5. We predict the hosts using both majority vote and the best alignment for BLASTN. The majority vote strategy assigns the most common alignment host to the virus. The strategy of the best alignment assigns the label using the host with the best alignment. The latter yielded better performance and we thus reported the results in Fig. \ref{fig:figure6}. The performance of majority vote can be found in Additional file 1: Fig. S1.

We compared the performance of HostG with other virus host classification tools in Fig. \ref{fig:figure6}. To ensure a fair comparison, we retrained vHULK and RaFAH using our training data. Other learning models are either hard to retrain or were previously trained using similar training data as ours and thus we directly tested them using the pre-trained models.  For alignment-based method VPF-Class, we directly used their database and run it on the TEST dataset.
Fig. \ref{fig:figure6} shows that HostG outperforms other pipelines across different ranking. With the increase of the ranking, the performance of all pipelines increases. This is expected because the higher taxonomic ranking has more relationship data to learn. In addition, features of higher-ranking taxonomic groups tend to be more distinctive. The results show that HostG achieves both high prediction accuracy and prediction rate. Although the performance of BLASTN in Fig. \ref{fig:figure6} is better than some of the learning-based pipelines, BLASTN can only return predictions for 65.5\% of the viruses in the TEST dataset. All other methods predicted the hosts for more than 90\% of the viruses. We also recorded the prediction results using provided parameters of RaFAH and vHULK in Additional file 1: Fig. S2. The results are much better than Fig. \ref{fig:figure6}. This is likely caused by the overlap between the TEST dataset and the data used for training the latest RaFAH and vHULK models.

\begin{figure}[h!]
    \centering
    \includegraphics[width=0.65\linewidth]{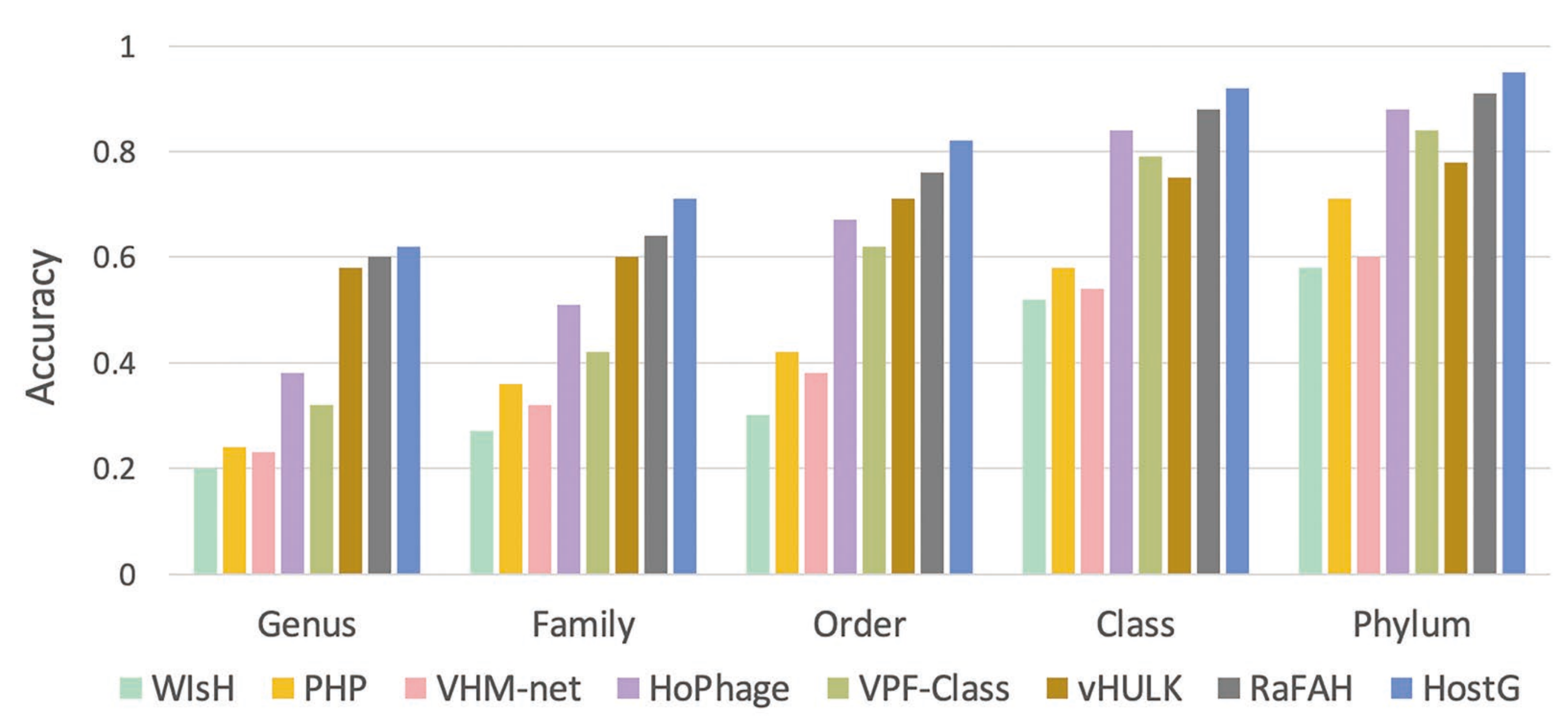}
    \caption{Host prediction accuracy for contigs without alignment results. X-axis: taxonomic rankings. Y-axis: accuracy.}
    \label{fig:figure8}
\end{figure}

Then, we further investigated the performance of the learning-based models for viruses that lack significant alignments with reference prokaryotic genomes. In this experiment, only viruses without BLASTN alignments will be used as test sequences. The results in Fig. \ref{fig:figure8} reveal that HostG still renders the best performance even when there are no statistically significant alignments between query virus and host. We also evaluated how the similarity between testing sequences and training sequences affects the prediction performance. The performance shown in Additional file 1: Fig. S2 reveals that sequence with higher similarity will achieve a better accuracy

\subsection{Improvement of GCN with ECE}

\begin{figure}[h!]
    \centering
    \includegraphics[width=0.7\linewidth]{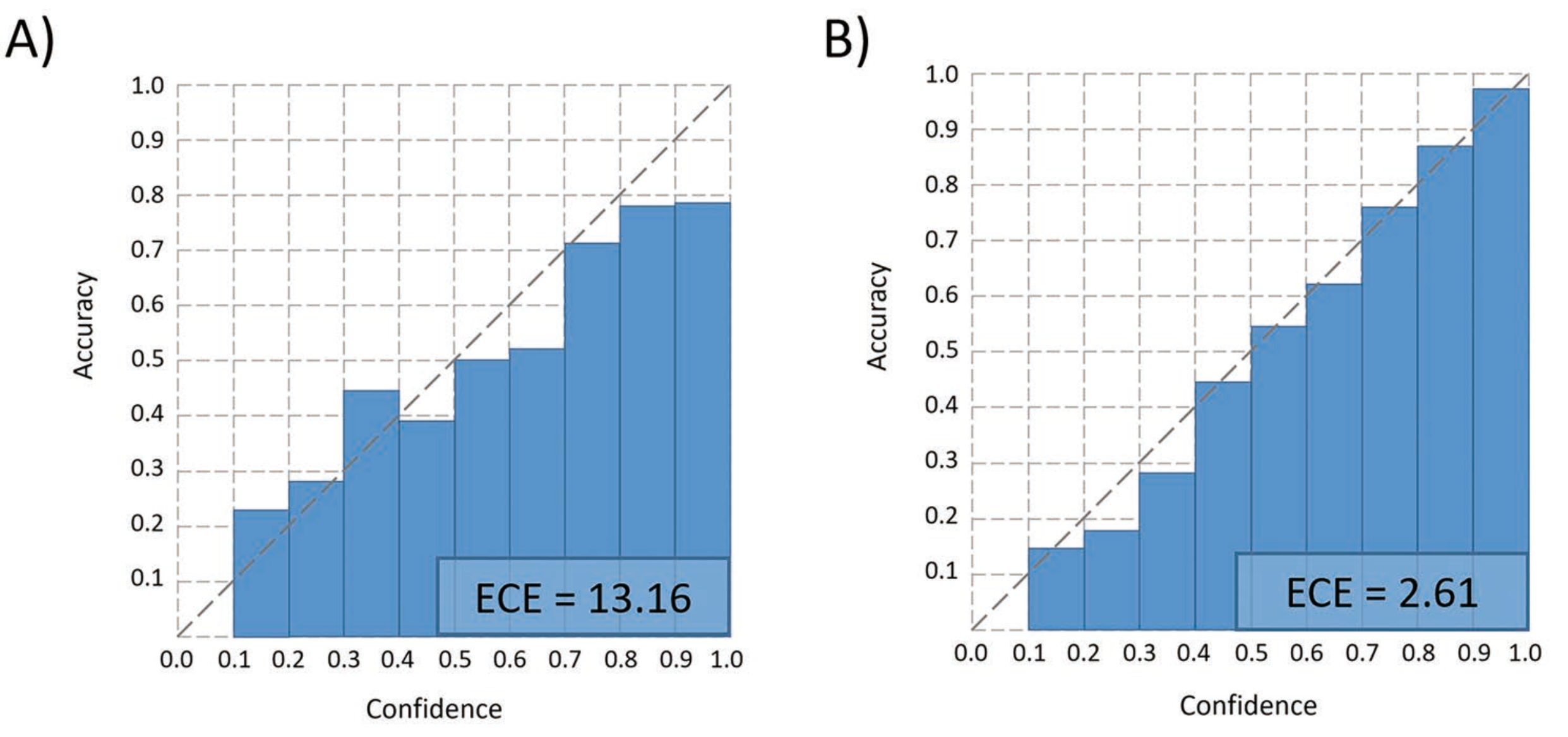}
    \caption{Accuracy vs. confidence (SoftMax value) before and after adding the ECE loss at the order level. ECE decreases from 13.16 to 2.61. X-axis: confidence (SoftMax value). Y-axis: accuracy.}
    \label{fig:figure9}
\end{figure}

As described in Section \ref{sec:ece}, we combine ECE and L2 to update the parameters in GCN. We divide the SoftMax value into 10 bins, so each bin covers a region of size 0.1. Fig. \ref{fig:figure9} shows the results before (Fig. \ref{fig:figure9}(A)) and after (Fig. \ref{fig:figure9}(B)) adding ECE in the training process. After adding ECE to the objective function, the bin with higher confidence (SoftMax value) has higher accuracy. For example, The number of samples with SoftMax values above 0.8 is 76, which corresponds to the accuracy of 95.7\% and prediction rate of 56\%. Thus, adding ECE to L2 allows us to achieve higher accuracy with a sacrifice of prediction rate. Users can adjust the threshold of confidence (SoftMax value) according to their needs.

\begin{figure}[h!]
    \centering
    \includegraphics[width=0.65\linewidth]{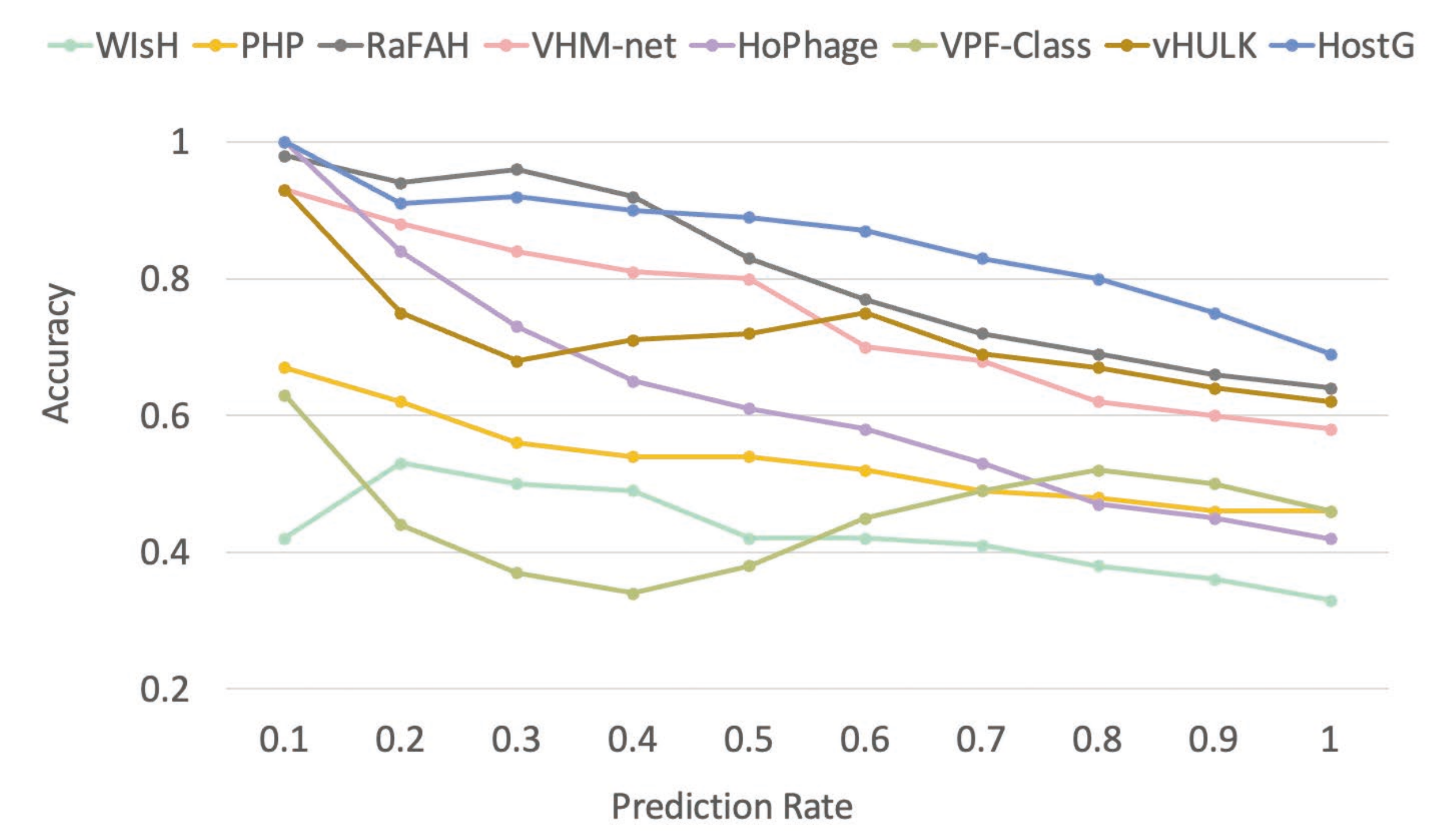}
    \caption{Comparison of the accuracy and prediction rate on the learning-based tools at the rank of genus. Each data point on a line corresponds to a different confidence threshold. X-axis: prediction rate. Y-axis: accuracy.}
    \label{fig:figure10}
\end{figure}

Other learning-based tools also provide a score for ranking their predictions. We first sorted the prediction according to the SoftMax value (or the score provided by other tools) and then showed the comparison at genus, family and order level in Fig. \ref{fig:figure10} and Additional file 1: Fig. S3. As expected, the accuracy and tends to decrease with the increase of the prediction rate. Fig. \ref{fig:figure10} and Additional file 1: Fig. S3 indicate that HostG can achieve higher host prediction accuracy than most of the existing tools under the same prediction rate across different taxonomic ranking. In addition, HostG achieves 100\% accuracy at the order, family, and genus level when the SoftMax thresholds are 0.88, 0.89, and 0.94, respectively. We also recorded the F1-scores of different tools in Additional file 1: Fig. S5. The result shows that HostG can achieve higher F1-score across different prediction rates.

\subsection{Performance on short contigs}

\begin{figure}[h!]
    \centering
    \includegraphics[width=0.65\linewidth]{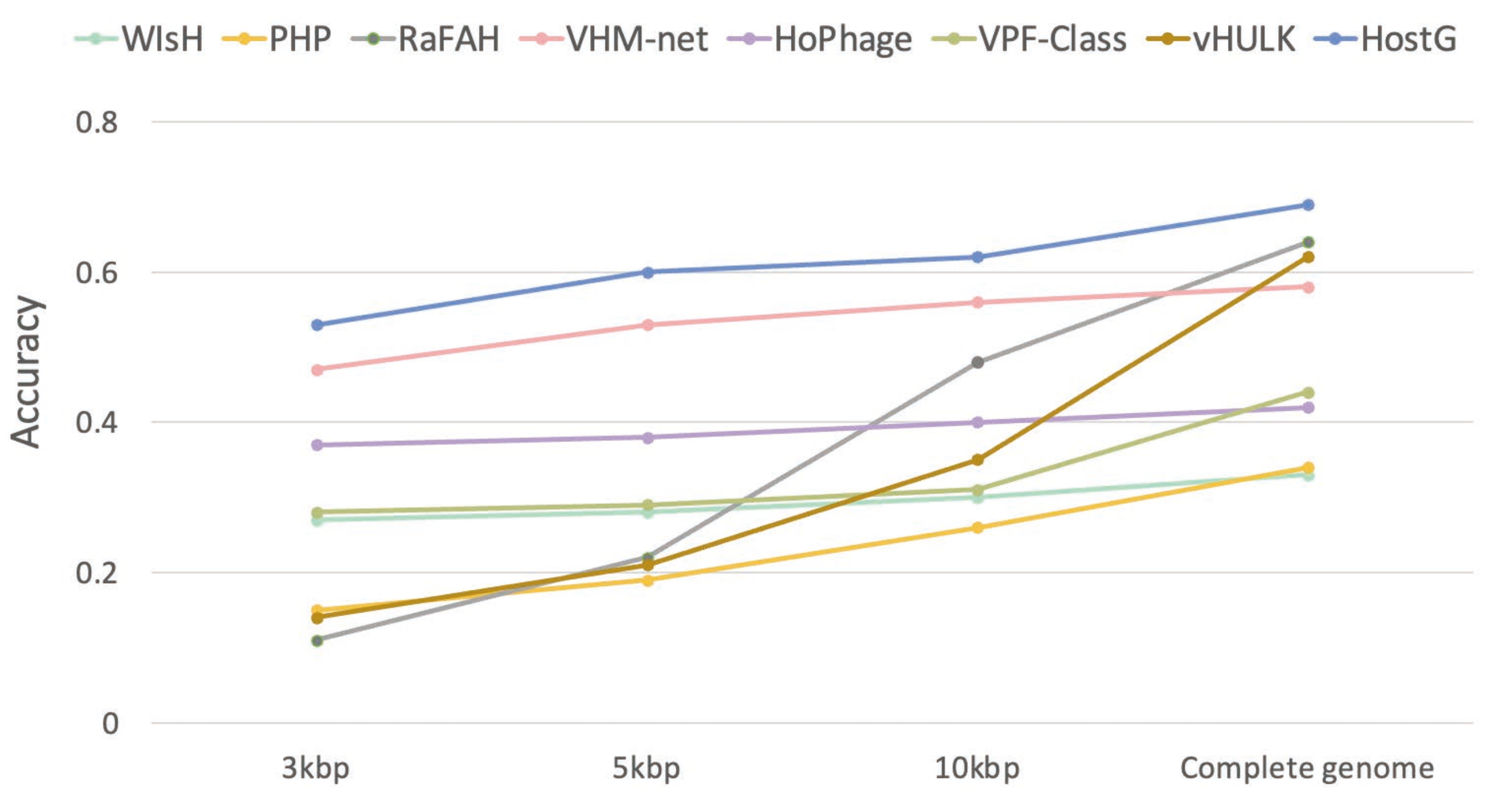}
    \caption{Prediction performance on short contigs. X-axis: length of the input contigs. Y-axis: accuracy.}
    \label{fig:figure11}
\end{figure}

While the previous experiments were conducted using whole genomes, we will investigate how the length of input contigs influences the prediction performance. First, following the experimental setting in Section 3.2, we randomly selected a start position and sampled contigs in three different lengths (3kbp, 5kbp, 10kbp) from the viral genomes in the TEST dataset. Then we ran all the pipelines and recorded the predictions. As shown in Fig. \ref{fig:figure11}, although the performance of the all methods decreases with the decrease of the contigs’ length, HostG still outperforms the state-of-the-art methods at three taxonomic rankings. 

\begin{figure}[h!]
    \centering
    \includegraphics[width=0.7\linewidth]{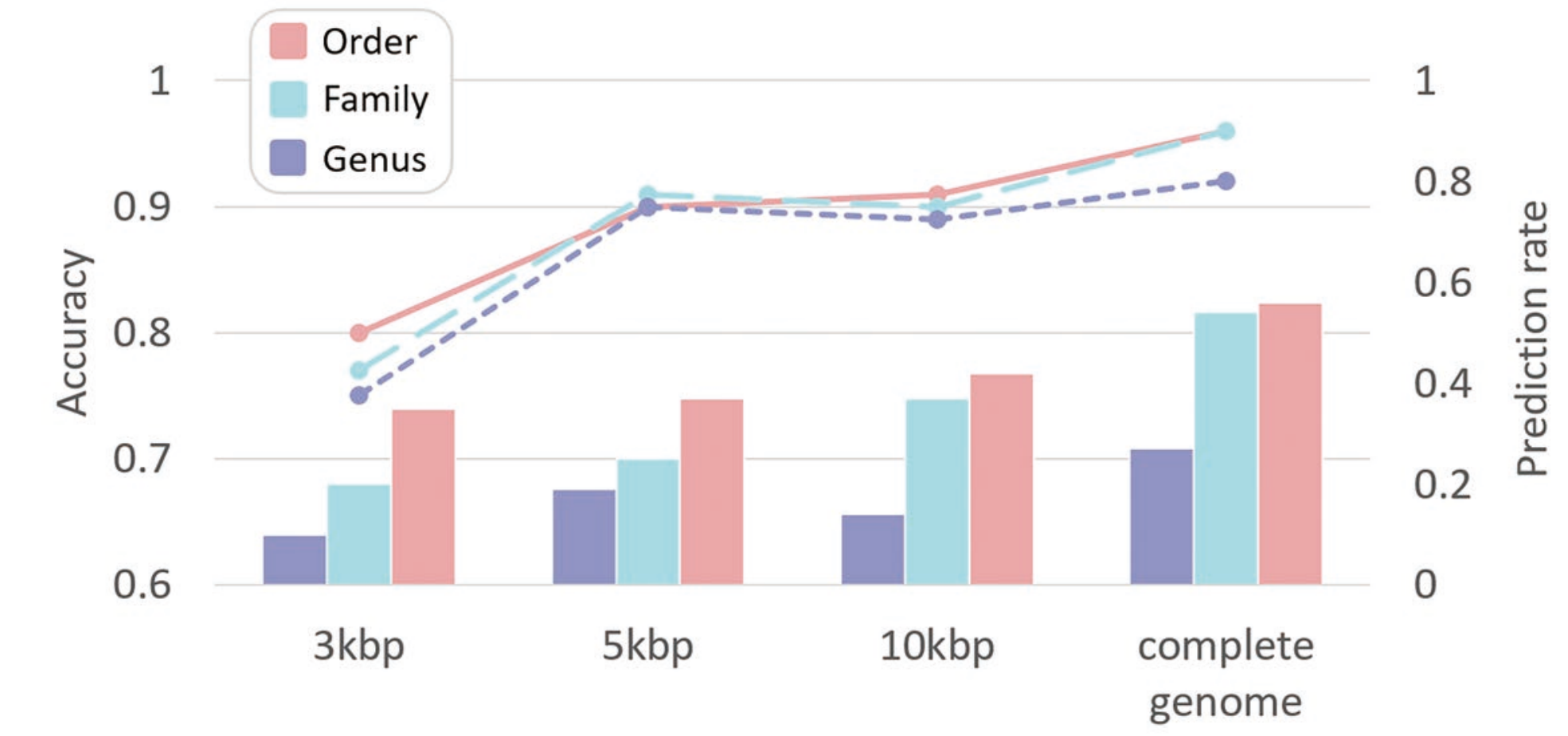}
    \caption{Prediction performance on short contigs with SoftMax threshold above 0.8. Line-plot: the accuracy vs. length of contigs. Bar-plot: the prediction rate vs. length of contigs.}
    \label{fig:figure12}
\end{figure}

Fig. \ref{fig:figure12} shows the classification performance of HostG with a SoftMax threshold above 0.8. Although there is a sacrifice of prediction rate, the predicted labels become more accurate for the short contigs. The results suggest that HostG is still reliable for short inputs when users specify a stringent SoftMax cutoff.

\subsection{Extension to hosts with new taxonomic labels}
To test the performance of HostG on predicting the hosts from new taxa, we designed two experiments using the 139 new virus-host pairs obtained by single-cell viral tagging \cite{dvzunkova2019defining}. In this dataset, the genus labels of the host genomes are new compared to the 1,426 virus-host relationships in the VHM dataset. Thus, lacking training samples on these new labels prevents supervised learning models such as CNN from predicting the correct labels for the 139 new viruses. However, HostG can conveniently include hosts from new taxa by adding the corresponding nodes in the knowledge graph. Then, the new taxonomic label can be propagated by edges during training.

\begin{figure}[h!]
    \centering
    \includegraphics[width=0.9\linewidth]{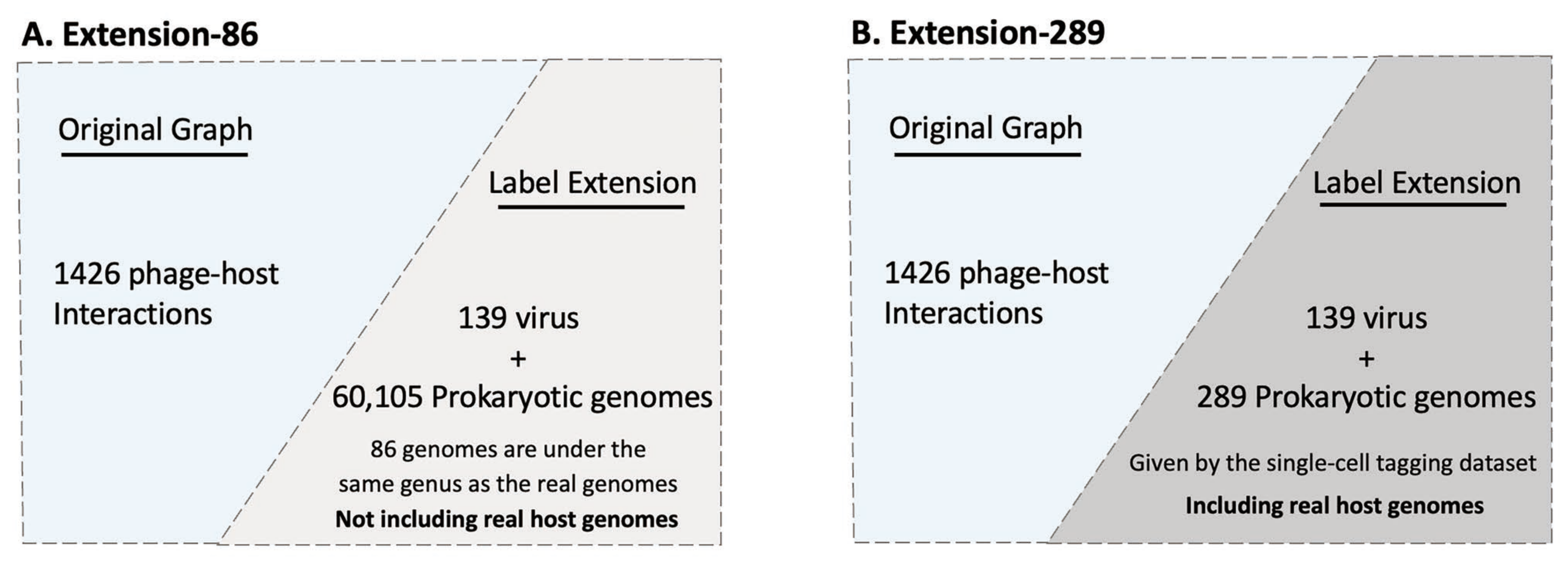}
    \caption{Two methods to extend the knowledge graph for new host labels. (A) Graph extension by adding 60,105 prokaryotic genomes and 139 query viruses. (B) Graph extension by adding 289 prokaryotic genomes in the single-cell tagging dataset and 139 query viruses.}
    \label{fig:extension}
\end{figure}

As shown in Fig. \ref{fig:extension}, we considered two scenarios that can benefit from label extension. In Fig. \ref{fig:extension} (A), the user lacks specific information about the hosts of some query viruses and thus, add nodes for all 60,105 prokaryotic genomes obtained from the NCBI genome database (before 2020) to extend the knowledge graph. Of the 60k$+$ genomes, 86 genomes have the same genus label as the real host genomes. Thus, as mentioned in \ref{sec:extend}, the genus label of the real hosts can be integrated into the original graph. To add the difficulty, we also removed the real host genomes to test whether the model can predict the hosts' genus label when the real host genomes are not included. Fig. \ref{fig:extension} (B) focuses on the second scenario where the user has access to the real host genomes, such as those assembled from the same type of environmental samples. So nodes of 289 prokaryotic genomes given by the single-cell tagging dataset are added to the graph.

\begin{figure}[h!]
    \centering
    \includegraphics[width=0.65\linewidth]{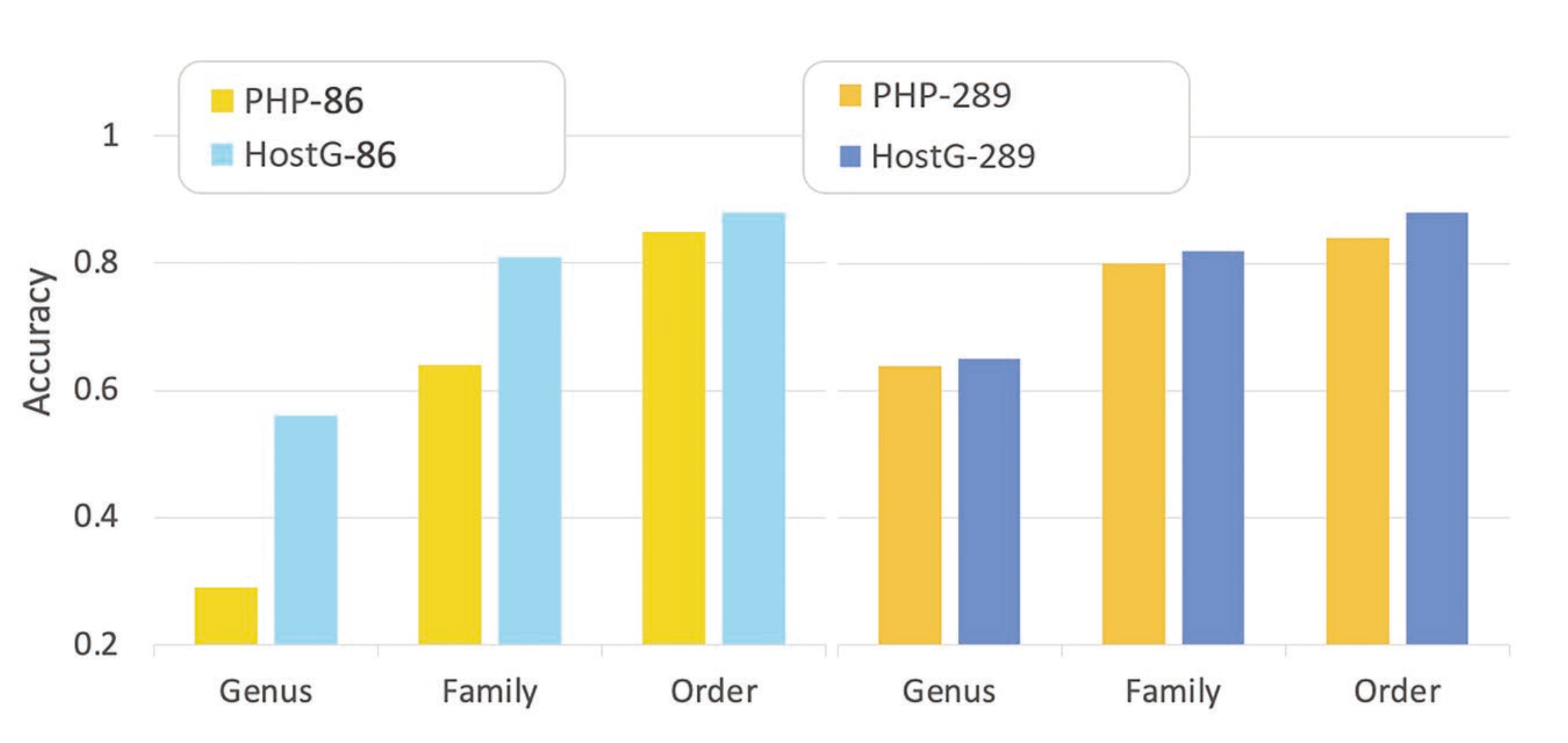}
    \caption{Prediction performance on the single-cell viral tagging dataset. ``-86'': trained and predicted on extension-86 shown in Figure \ref{fig:extension} (A). ``-289'': trained and predicted on extension-289 shown in Figure \ref{fig:extension} (B).}
    \label{fig:figure13}
\end{figure}

Fig. \ref{fig:figure13} shows the results of HostG trained on the extended knowledge graphs. Because PHP supports model retraining for label extension even when the training set does not contain the labels of the host species, we compared the accuracy with the outputs of PHP. As shown in Fig. \ref{fig:figure13}, the extended version of HostG can achieve higher accuracy in both cases. As expected, both HostG and PHP have better performance when the actual host genomes are used as the labeled sequences, which is expected. When the actual host genomes are not in the knowledge graph, HostG can still utilize the prokaryotes in the same taxa to make more reliable predictions than PHP. 

\begin{figure}[h!]
    \centering
    \includegraphics[width=0.7\linewidth]{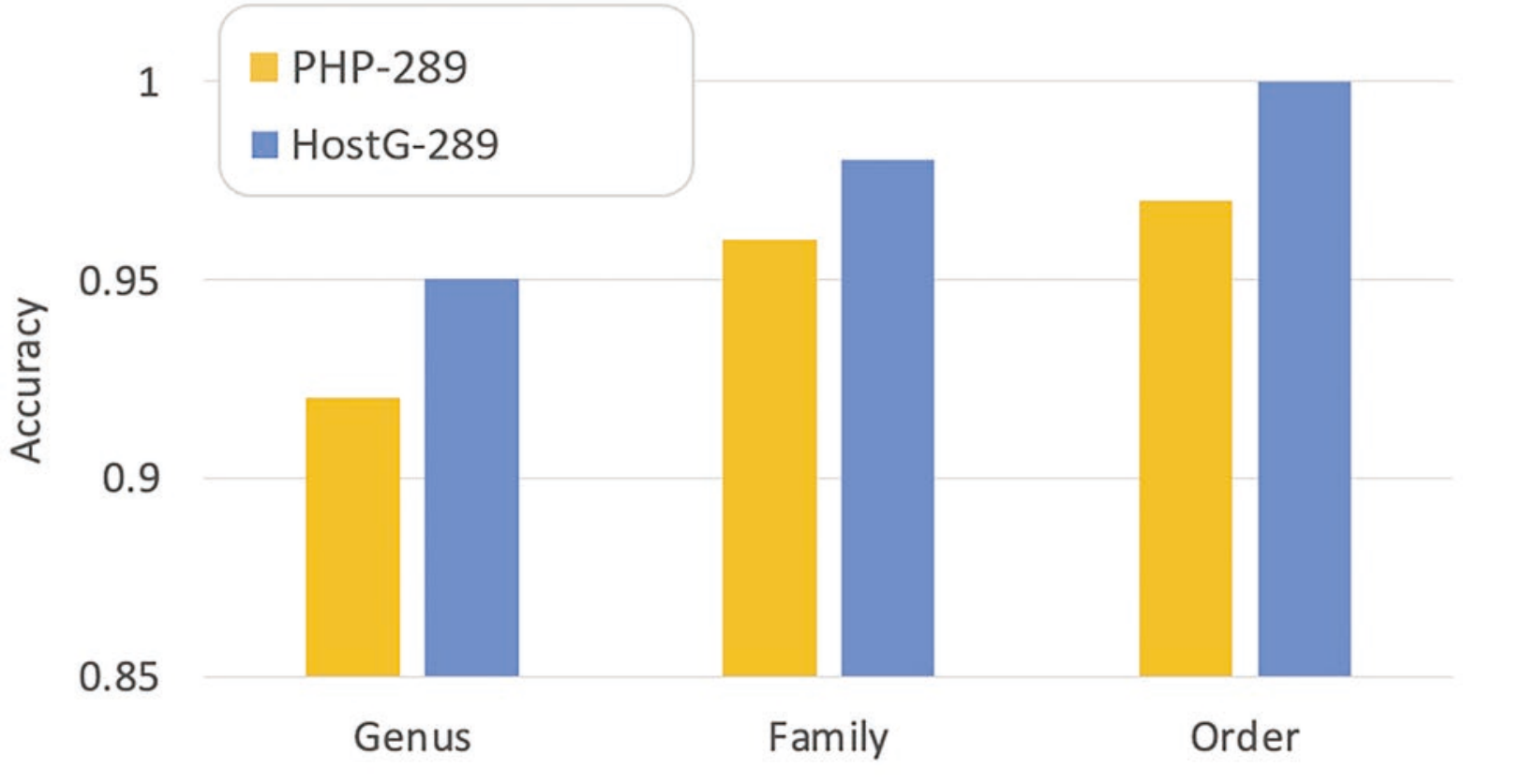}
    \caption{Prediction accuracy for contigs with the highest 20\% SoftMax values (or scores). X-axis: taxonomic ranking. Y-axis: accuracy. }
    \label{fig:figure14}
\end{figure}

We also recorded the results of HostG with the highest 20\% SoftMax values and PHP with the highest 20\% scores. As shown in Fig. \ref{fig:figure14}, imposing the thresholds renders higher accuracy. \

\vspace{-0.2cm}
\section{Discussion}

As shown in the experiments, the performance of alignment-based methods heavily rely on the reference database. The ambiguous hits or lack of shared regions with host genomes can decrease the classification accuracy and the prediction rate. Existing learning-based tools like PHP cannot achieve good performance at low taxonomic ranking, such as genus and family. The results become even worse when the query contigs are short. In this work, we demonstrated that HostG outperforms the state-of-the-art methods for host prediction. Rather than only using the DNA patterns from virus-host pairs, we also consider the protein similarity between viruses to construct the knowledge graph. Then, the semi-supervised learning method, GCN, enables HostG to exploit features from both labeled and unlabeled nodes in the knowledge graph and predict hosts for query viruses. To ensure the reliability of HostG, we employed ECE to calibrate the confidence of the predictions so that users can achieve higher accuracy by setting a threshold according to their needs. Finally, we demonstrated that HostG can predict new taxonomic labels through the extension capability of the knowledge graph.

Although HostG has greatly improved host prediction, we have several goals to optimize in our future work. First, the length of the contigs will influence the classification performance. In order to improve the accuracy of the short contigs, we will investigate whether more biological features can be incorporated in the knowledge graph construction. Second, as shown in Additional file 1: Table. S2 in the supplementary file, HostG a has longer running time than some tools. The bottleneck of HostG is the calculation of the alignment similarities. We will explore whether the alignment can be replaced by a more efficient method to save computational resources. 

\section{Conclusions}
In this work, we present a semi-supervised learning model, named HostG, to conduct host prediction for novel viruses. We tested HostG on both simulated and real sequencing data and the results demonstrated that it outperforms the state-of-the-art pipelines. This work will help to identify virus-host interactions in metagenomic data and will extend our understanding of newly identified viruses

\section*{Acknowledgements}
The computation was conducted at HPCC of City University of Hong Kong. 

\section*{Funding}
This work was supported by the Hong Kong Innovation and Technology Commission and City University of Hong Kong (Project 7005453) and HKIDS (9360163).

\subsection*{Availability of data and materials}
The source code of HostG is available at: \url{https://github.com/KennthShang/HostG}.\\
The training set, testing set, and the single cell viral tagging dataset are from: \url{https://github.com/congyulu-bioinfo/PHP/tree/master/virus-hostInteractionData}. The training set is listed in VHM\_PAIR\_TAX.xls. The testing set is listed in TEST\_PAIR\_TAX.xls. The single cell viral tagging dataset is detailed in PRJNA492716\_dataset.xls. They are also available via: \url{https://github.com/KennthShang/HostG/dataset}.

\bibliographystyle{unsrt}  
\bibliography{references}  

\end{document}